\documentclass[aps,prx,superscriptaddress,twocolumn,10pt,longbibliography]{revtex4-1}
\usepackage{times,color,amsmath,amsfonts,amsthm,graphicx,xcolor}
\usepackage{times,color,amsmath,amsfonts,amsthm,graphicx,xcolor,braket}
\usepackage{hyperref}
\hypersetup{
    colorlinks=true,
    breaklinks=true,
    citecolor=blue,
    linkcolor=blue,    
    urlcolor=blue,
}

\usepackage{breakurl}
\hypersetup{breaklinks=true}
\usepackage[version=3]{mhchem}
\usepackage{subfigure}
\usepackage{ulem}
\usepackage[super]{nth}
\usepackage{soul}

\begin{document}

\title{Ground-state properties of the hydrogen chain:\\
dimerization, insulator-to-metal transition, and magnetic phases}

\author{Mario Motta}
\thanks{These authors contributed equally to this work}
\affiliation{IBM Almaden Research Center, San Jose, CA 95120, USA}
\affiliation{Division of Chemistry and Chemical Engineering, California Institute of Technology, Pasadena, CA 91125, USA}

\author{Claudio Genovese}
\thanks{These authors contributed equally to this work}
\affiliation{SISSA -- International School for Advanced Studies, Via Bonomea 265, 34136 Trieste, Italy}

\author{Fengjie Ma}
\thanks{These authors contributed equally to this work}
\affiliation{The Center for Advanced Quantum Studies and Department of Physics, Beijing Normal University, Beijing, Beijing 100875, China}

\author{Zhi-Hao Cui}
\thanks{These authors contributed equally to this work}
\affiliation{Division of Chemistry and Chemical Engineering, California Institute of Technology, Pasadena, CA 91125, USA}

\author{Randy Sawaya}
\thanks{These authors contributed equally to this work}
\affiliation{Department of Physics and Astronomy, University of California, Irvine, CA 92697-4575, USA}

\author{Garnet Kin-Lic Chan}
\affiliation{Division of Chemistry and Chemical Engineering, California Institute of Technology, Pasadena, CA 91125, USA}

\author{Natalia Chepiga}
\affiliation{Institute for Theoretical Physics, University of Amsterdam, Science Park 904 Postbus 94485, 1090 GL Amsterdam, The Netherlands}

\author{Phillip Helms}
\affiliation{Division of Chemistry and Chemical Engineering, California Institute of Technology, Pasadena, CA 91125, USA}

\author{Carlos Jim\'enez-Hoyos}
\affiliation{Department of Chemistry, Wesleyan University, Middletown, CT 06459, United States}

\author{Andrew J. Millis}
\affiliation{Center for Computational Quantum Physics, Flatiron Institute, New York, NY 10010, USA}
\affiliation{Department of Physics, Columbia University, New York, NY 10027, USA}

\author{Ushnish Ray}
\affiliation{Division of Chemistry and Chemical Engineering, California Institute of Technology, Pasadena, CA 91125, USA}

\author{Enrico Ronca}
\affiliation{Max Planck Institute for the Structure and Dynamics of Matter, Luruper Chaussee 149, 22761 Hamburg, Germany}
\affiliation{Istituto per i Processi Chimico Fisici del CNR (IPCF-CNR), Via G. Moruzzi 1, 56124 Pisa, Italy}

\author{Hao Shi}
\affiliation{Center for Computational Quantum Physics, Flatiron Institute, New York, NY 10010, USA}

\author{Sandro Sorella}
\affiliation{SISSA -- International School for Advanced Studies, Via Bonomea 265, 34136 Trieste, Italy}
\affiliation{Democritos Simulation Center CNR--IOM Istituto Officina dei Materiali, Via Bonomea 265, 34136 Trieste, Italy}

\author{Edwin M. Stoudenmire}
\affiliation{Center for Computational Quantum Physics, Flatiron Institute, New York, NY 10010, USA}

\author{Steven R. White}
\affiliation{Department of Physics and Astronomy, University of California, Irvine, CA 92697-4575, USA}

\author{Shiwei Zhang}
\thanks{szhang@flatironinstitute.org}
\affiliation{Center for Computational Quantum Physics, Flatiron Institute, New York, NY 10010, USA}
\affiliation{Department of Physics, College of William and Mary, Williamsburg, VA 23187-8795, USA}

\collaboration{Simons collaboration on the many-electron problem}

\begin{abstract}
 Accurate and predictive computations of the quantum-mechanical behavior of many interacting electrons
in realistic atomic environments
are critical for the theoretical design of materials with desired properties,
and require solving the grand-challenge problem of the many-electron Schr\"{o}dinger equation.
An infinite chain of equispaced hydrogen atoms is perhaps the simplest realistic model for a bulk material, 
embodying several central themes of modern condensed matter physics and chemistry, 
while retaining a connection to the paradigmatic Hubbard model.
Here we report a combined application of cutting-edge computational methods to determine 
the properties of the hydrogen chain in its quantum-mechanical ground state.
Varying the separation between the nuclei leads to a rich phase diagram, 
including a Mott phase with quasi long-range antiferromagnetic order, electron density dimerization with power-law correlations, an insulator-to-metal transition 
and an intricate set of intertwined  magnetic orders. 
\end{abstract}

\maketitle

\section{Introduction}

The study of interacting many-electron systems  is profoundly important to many areas of science, including 
condensed-matter physics and quantum chemistry.
Materials properties result from a delicate interplay of competing factors including
atomic geometry and structure, quantum mechanical delocalization of electrons from atoms,
the entanglement implied by quantum statistics, and electron-electron interaction.
Capturing such effects accurately is essential for understanding  properties, for predictive computations, 
and for the realization of materials genome or materials-by-design initiatives, but requires a sufficiently accurate 
solution of  the interacting electron problem.

Over the years, many approaches to the problem of solving the many-electron Schr\"{o}dinger equation have been formulated. 
Density functional theory (DFT) obtains the ground state energy of the many-body system in terms of the solution of an auxiliary one electron 
problem plus a self-consistency condition. DFT based methods have had enormous impact on materials science and condensed matter physics,  
but the approaches become less reliable, or even break down, in the presence of strong electronic correlation effects including magnetic, 
structural, conductive and superconductive phase transitions \cite{mott1949basis,mott1961transition,Mott_RMP_1968,Dagotto_RMP_1994,Georges_RMP_1996,Imada_RMP_1998,Lee_RMP_2006}. 
Alternative approaches in terms of model systems such as the single-orbital Hubbard model have contributed very significantly to our 
understanding of electronic correlation physics, but involve dramatic simplifications, including neglect of many of the terms in the 
physical Coulomb interaction and truncation to a small fixed orbital basis, whose consequences are not fully understood.   
The need to establish systematic approaches that are chemically realistic
and fundamentally many-body is thus intensely investigated.

A linear chain of hydrogen atoms 
($N$ protons equispaced along a line, with $N$ electrons)  \cite{Hachmann_JCP_2006,Alsaidi2007,Sinitski_JCP_2010,Sorella_PRB_2011,Motta_PRX_2017}
embodies many central themes of modern condensed matter physics,
 while retaining a certain degree of computational tractability.
It features a periodic atomic potential, incorporates realistic Coulomb interactions, requires the construction of accurate and compact basis sets,
and yet maintains a connection with the fundamental Hubbard model which has been a hallmark of the theory of interacting fermions.

Here we deploy a combination of the most advanced and accurate many-body methods to investigate the ground-state phase diagram of the H chain.
%
Our methods have been benchmarked in a previous 
collaborative effort to obtain the ground-state equation of state 
\cite{Motta_PRX_2017}. In this work, we have introduced methodological advances
to compute properties beyond the total energy. 
There have also been several previous  
studies of this system \cite{Hachmann_JCP_2006,Alsaidi2007,Sinitski_JCP_2010,Sorella_PRB_2011}. However, 
they were restricted either to
small basis sets or finite system sizes, which prevented a realistic 
description of the H chain, or by their 
accuracy and capabilities, which prevented reliable resolution of the 
delicate scales and discovery of all the phases.
The synergistic application of complementary methods 
distinguishes the present work 
and allows us 
to make robust predictions with a multi-messenger approach 
in this challenging problem. 

\section{Methods}

We consider a system of $N$ protons at fixed, equally spaced positions along a line, with $N$ electrons:
\begin{equation}
\label{eq:ham1st}
\begin{split}
\hat H = &-\frac{1}{2} \, \sum_{i=1}^N \nabla^2_i + \sum_{i<j=1}^N \frac{1}{|{\bf
{r}}_i - {\bf{r}}_j|} \\
            &- \sum_{i,a=1}^N \frac{1}{|{\bf{r}}_i - {\bf{R}}_a|} + \sum_{a<b=1}^N \frac{1}{|{\bf{R}}_a - {\bf{R}}_b|} \,,\\
\end{split}
\end{equation}
where $({\bf{r}}_1 \dots {\bf{r}}_N)$ are the electron coordinates, 
and ${\bf R}_a=aR {\bf e_z}$ is the coordinate of the $a$-th proton. 
This Hamiltonian has been studied in finite basis or finite systems (or both) and has generated increasing interest \cite{Hachmann_JCP_2006,Sinitski_JCP_2010,Sorella_PRB_2011,Zhao_JCTC_2017,Kong_JP_2017,Zheng_FP_2018,ronca2017time}.
We use atomic units throughout, in which energies are measured in Hartrees ($me^4/\hbar^2$) and lengths in units of the Bohr radius $a_B=\hbar^2/(me^2)$. 
In the thermodynamic limit (TDL) of infinite system size at zero temperature, which is our primary focus, our system is characterized by only one parameter,  $R$.


We access the ground-state properties using multiple first-principles many-body methods, including variational Monte Carlo (VMC), 
standard and sliced-basis density-matrix renormalization group (DMRG, sb-DMRG) \cite{White_PRL_1992,white1999ab,sharma2012spin,Stoudenmire_PRL_2017},
auxiliary-field (AFQMC) \cite{Zhang_PRL_2003} and lattice-regularized diffusion Monte Carlo (DMC) \cite{Casula_PRL_2005}.  
For reference, independent-electron calculations are also performed, including restricted (RHF) and unrestricted (UHF) Hartree-Fock \cite{Szabo_book_1982} 
and DFT \cite{Martin_book_2004}. 
While it is not practically possible with any single method to simultaneously converge all aspects of the electronic structure calculations 
(such as basis set limit, many-electron correlation, and thermodynamic limit) across different regimes of the ground-state phase diagram, 
we draw our conclusions based on the convergence of multiple approaches to a consistent physical picture.

In this work, we introduce several technical advances which enabled the calculations on the hydrogen chain presented here.
Within AFQMC, we implemented techniques for the study of metal-insulator transitions
based on the modern theory of polarization, and the calculation of structure factors in
chemical simulations. A procedure was developed for generating basis sets for confined systems (as in the planes perpendicular to the chain)
from downfolded virtual Kohn-Sham orbitals \cite{Ma_PRL_2015}. 
Within DMC, we have implemented a scheme for computing unbiased estimates of correlation functions and developed
basis sets for the short-$R$ regime, as well as a more efficient parametrization of the trial wave function ansatz.
 Within DMRG, we introduced new basis set techniques building on the idea of Gausslets \cite{White_PRB_2019},
and implemented the computation of entanglement entropy and dimerization in electronic density.
To approach the continuum 
limit 
for small values of $R$, we devised and extensively
compared basis sets 
including planewaves, continuum coordinate-space, and problem-specific Gaussians with diffuse and polarized functions, 
as well as the new basis sets mentioned above.
To approach the TDL in our numerical computations, we study increasingly long chains with open boundary conditions (OBC),
or supercells with periodic Born-von-Karman boundary conditions (PBC).
Further descriptions of our methods and discussions of technical advances and issues 
can be found in the supplemental information (SI).

\section{Results}

In this section, we present our results as a function of $R$. We start by describing the large-$R$
regime, where we find that the chain features quasi-long-range antiferromagnetic correlations, followed by dimerization
in several properties, namely electronic density, kinetic energy and entanglement entropy as $R$ is reduced.
We then proceed to describe the short-$R$ regime, distinguished 
by a metal-insulator transition occurring at a critical bondlength $R_{\mathrm{MIT}}$, so that the chain is insulating for $R > R_{\mathrm{MIT}}$ and metallic for $R<R_{\mathrm{MIT}}$.
We then discuss the physical properties of the metallic phase, and show that the MIT 
arises from a self-doping mechanism.
 
\subsection{Large-$R$ regime}

\begin{figure}[t!]
\centering
\includegraphics[width=0.9\columnwidth]{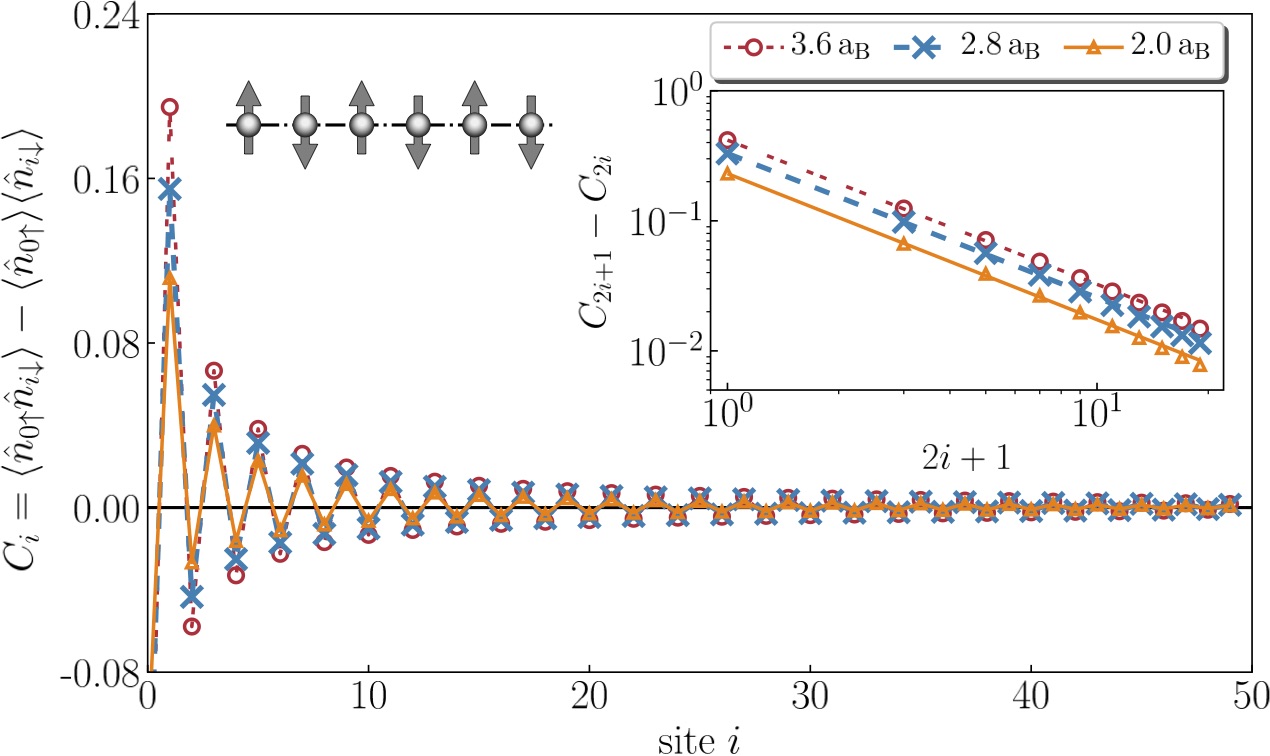}
\caption{(color online) {\bf{Insulating phase -- antiferromagnetic correlations.}}
Main: correlation function $C_i$ at $R=3.6$, $2.8$, and $2.0\,\mathrm{a_B}$ (red circles, blue crosses, and orange triangles)
computed with DMRG for a chain of $N=50$ with OBC.
The oscillations, with wavelength $\lambda = 2R$, show AFM correlations. 
Inset: oscillation amplitudes $C_{2i+1}-C_{2i}$ versus $i$ (log-log scale). Lines show the results of a linear fit.}
\label{fig:f1}
\end{figure}

\subsubsection{Antiferromagnetic correlations}

We begin our study of the H chain phase diagram at large proton-proton separation.
Here to a first approximation the system is a collection of isolated H atoms, each  with a single electron in the atomic $1s$ orbital.
This is very similar to the half-filled  Hubbard model in the large coupling ($U/t$) limit.
However the weakly bound nature of H$^-$ and its very diffuse orbitals can create excitons with strong binding in the H chain. 

At large $R$, the correlations in the H chain can be characterized in terms of a spin-$\frac{1}{2}$ Heisenberg chain.
The Heisenberg chain is a critical system, with power-law decay of AFM spin-spin correlations,
$\langle \hat{\bf{S}}_0 \cdot \hat{\bf{S}}_i\rangle$.
In response to a local perturbation (such as a local magnetic field), 
one observes local ordering (such as local N\'eel order) which decays as a power law in the distance from the perturbation. 
To probe AFM correlations in the H chain, in Fig.~\ref{fig:f1}
we present the quantity $C_i = \langle \hat{n}_{0\uparrow} \hat{n}_{i\downarrow} \rangle - 
\langle \hat{n}_{0\uparrow} \rangle \langle \hat{n}_{i\downarrow} \rangle$ computed with DMRG in the minimal (STO-6G) basis, 
where $\hat{n}_{i \sigma}$ denotes the number of electrons occupying (orthogonalized) atomic orbital $i$ with spin polarization $\sigma$ along $z$.
$C_i$ oscillates with wavelength $\lambda = 2R$, 
corresponding to the  N\'eel  vector of two sublattice antiferromagnetism; the wavevector may also be thought of as twice the Fermi 
wave-vector $q = \frac{2\pi}{\lambda} = 2\,k_F^0$ of a paramagnetic 1D ideal Fermi gas of density $\rho = \frac{1}{R}$.

In Fig.~\ref{fig:f1}, we further observe that the oscillations in $C_i$ decay with a power-law envelope, $C_i = C_0 i^{-\eta} \, (-1)^i$. 
The decrease of $C_0$ with $R$ (see inset) indicates weakening of AFM as the bondlength is reduced. The power-law decay of the AFM correlation 
is consistent with quasi long-range order in 1D.
The  fitted exponent of $\eta \simeq 1.11(1)$, likely affected by finite-size effects, is slightly higher than 
the prediction from conformal field theory, which gives $\langle\mathbf{S}_i \cdot \mathbf{S}_0\rangle \propto (-1)^i \sqrt{\ln i}/i^\eta$, 
with $\eta = 1$ 
for systems within the same universality class as the 1D Heisenberg chain  \cite{Affleck_JPA_1998,Affleck_Gepner_1989}.

\begin{figure}[!ht]
\centering
\includegraphics[width=0.9\columnwidth]{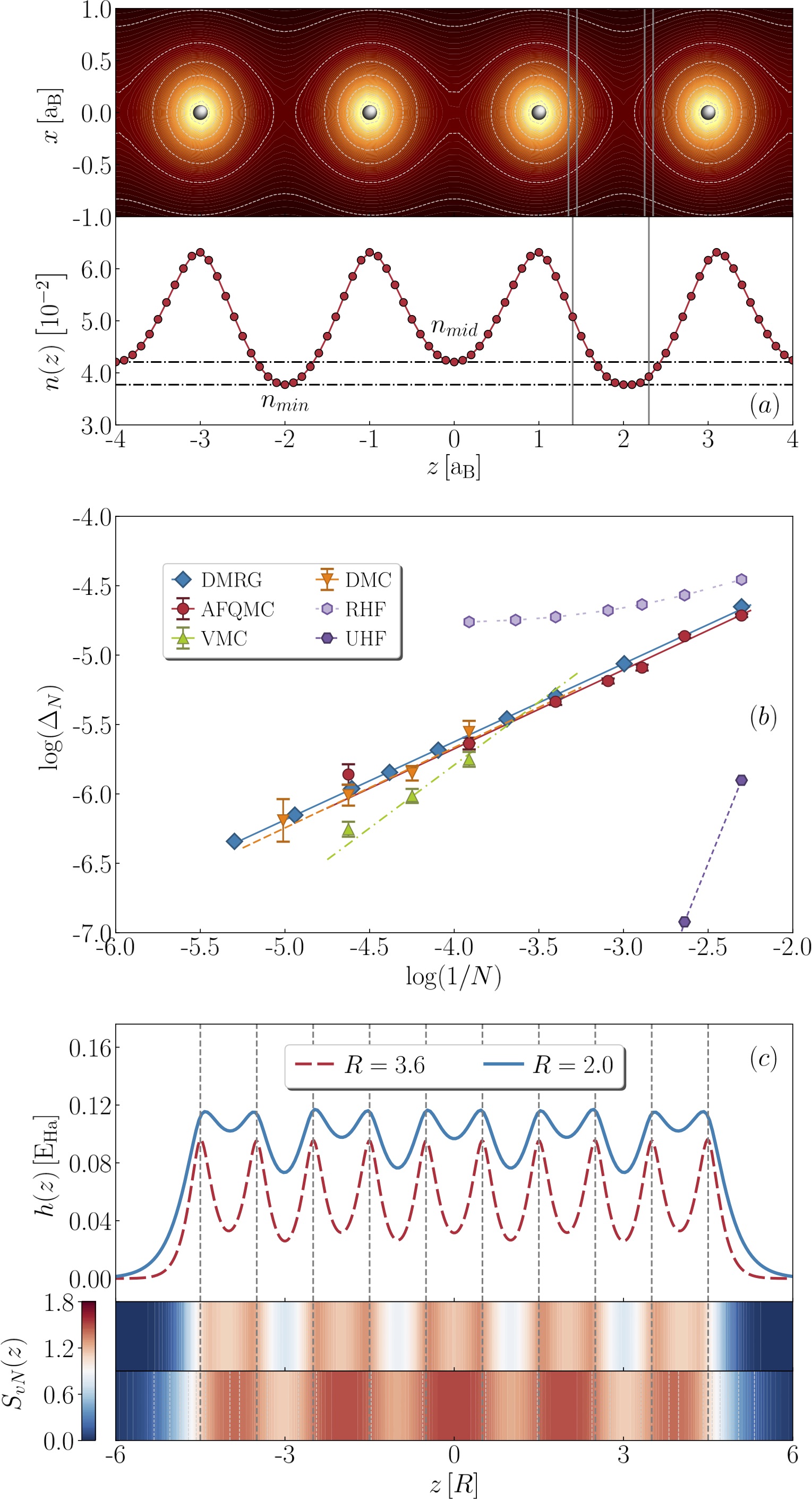}
\caption{(color online) {\bf{Insulating phase -- density dimerization.}}
(a) Top: electronic density $n(z)$ for $R=2.0\,\mathrm{a}_B$, with OBC.
Bottom: definition of the dimerization measure $\Delta_N$ for a system under OBC.
The gray vertical lines indicate how $n(z)$ is integrated along thin (width $\delta z=0.1\,\mathrm{a_B}$) slabs perpendicular 
to the chain.
(b) Dependence of dimerization measure $\Delta_N$ on the number of atoms in the open chain, $N$, from AFQMC, sb-DMRG, VMC, and DMC. 
$\Delta_N$ decays as a power-law, $\Delta_N \propto N^{-d}$ with exponent $d \simeq 0.5$ 
from correlated methods.
Results from HF are also shown for reference.
(c) Expectation value $h(z)$ of the kinetic energy operator (top) and von Neumann entropy $S_{vN}(z)$ (bottom)
from sb-DMRG, for a 10-atom chain with OBC, as a function of position $z$ along the chain axis for $R=3.6$ and $2.0\,\mathrm{a}_B$, respectively.
}
\label{fig:f2}
\end{figure}

\subsubsection{Dimerization}
We observe dimerization
as $R$ is reduced in the large $R$ regime. 
With PBC, dimerization can be probed by density-density correlations.
With OBC, or with a local perturbation, density dimerization can be measured by 
the electronic density, for example integrated along transverse slabs (i.e., over $x$ and $y$ for a finite $\delta z$),  $n(z)$.
The upper portion of Fig.~\ref{fig:f2}a shows $n(z)$ versus $z$, for a segment at the center of the chain under OBC, computed with AFQMC. 
The density has maxima at the proton sites and minima half way between. 
We define the dimerization measure for an $N$-atom chain, $\Delta_N$, by the difference between the two adjacent local minima of $n(z)$ in the center of chain
as illustrated in the lower portion of  Fig.~\ref{fig:f2}a.
When dimerization is present, we find that its amplitude increases as $R$ is decreased. 

In Fig.~\ref{fig:f2}b we investigate the dimerization in the H chain at $R=2.0\,\mathrm{a}_B$, using
AFQMC, sb-DMRG, DMC, and VMC. 
We find that, in the middle of the chain, dimerization decays 
with chain length as $\Delta_N \propto N^{-d}$,
with exponents $d = 0.57(3)$, $0.563(4)$, $0.58(11)$ and $0.90(26)$ 
from AFQMC, sb-DMRG, DMC and VMC, respectively.
The subtle power-law physics of 1D correlated systems is not easy to capture accurately with numerical methods.
It is encouraging that 
a quantitative agreement is seen 
between our very different many-body methods. 
HF is qualitatively incorrect here, giving either no order or true long-range order; DFT results are similar.  

It is illuminating to relate the dimerization order observed in the 
H chain to a kind of 
dimerization order which appears in the Heisenberg model, as seen in nearest-neighbor spin correlations,
$\langle (\hat{\bf{S}}_0 \cdot \hat{\bf{S}}_1) 
(\hat{\bf{S}}_i \cdot \hat{\bf{S}}_{i+1})\rangle$
under PBC.
In the case of OBC, the open ends act as a local perturbation 
which couples to the dimerization, making odd-numbered bonds 
stronger than even ones, with an amplitude that decays as a power 
law away from the ends.
Criticality can thus be observed by studying the strength of the dimerization in the center of a chain as a function of $N$, 
In the Heisenberg chain the dimerization profile as a function 
of site index $i$ is described by $\Delta_N(i)\propto\left[N\sin(\pi i/N)\right]^{-d}$, where $d = 1/2$.  
This is consistent with our observations for the H chain.
In the half-filled Hubbard model, we observe dimerization in the correlation function of the nearest-neighbor hopping, 
i.e., the kinetic energy operator, $\hat{h}_i = \sum_\sigma \big( \hat{a}^\dagger_{i,\sigma} \hat{a}_{i+1,\sigma} + \mathrm{h.c.} \big)$. 
As shown in the SI, the correlation function $H_i = \langle \hat{h}_0 (\hat{h}_i  - \hat{h}_{i+1}) \rangle$   
is found to again display a power law behavior  $ (-1)^i i^{-\eta'}$, with  $\eta' \sim 0.57(8)$.

Dimerization manifests itself not only in the electron density, but also in other observables such as the kinetic energy and entanglement entropy.
Results for a 10-atom H chain are shown in Fig.~\ref{fig:f2}c from sb-DMRG, where the kinetic energy is measured
as hopping (integrated in $x$ and $y$) between slices, and the entanglement entropy is measured across planes that bisect a bond.
At  $R=2.0\,\mathrm{a_B}$,  where dimerization is consistently  observed in larger systems from the electronic density,
$S_{vN}(z)$ shows five entangled dimers (bottom of lower panel).
The entanglement exhibits a striking asymmetry on the two sides of each proton along the chain, with a strength discrepancy 
much larger than in the electron density.

\begin{figure*}[t!]
\centering
\includegraphics[width=\textwidth]{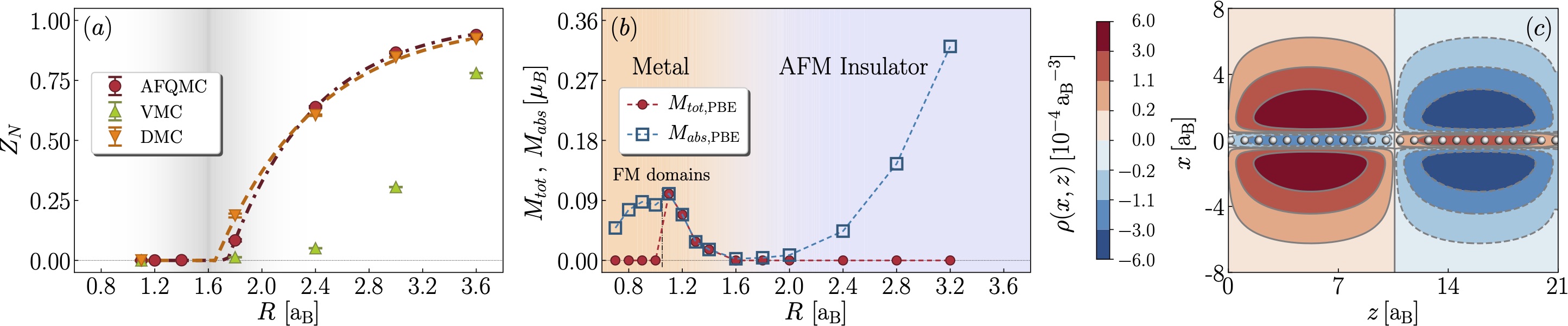}
\caption{
(color online) {\bf{Metal-insulator transition.}}
(a) Complex polarization measure as a function of $R$ from multiple methods,
identifying a MIT separating a region at smaller $R$ with 
$|Z_N|=0$ from $|Z_N|>0$ at larger $R$. 
$|Z_N|$ is plotted for $N=40$.
(b)
Magnetic phases from DFT-PBE, as indicated by $M_{tot}=\int{m(r) d^3r}/N$ and $M_{abs}=\int{\lvert m(r)\rvert d^3r}/N$, 
where $m(r)$ is the magnetization of the simulation cell. 
(c) Spin-density computed from DFT-PBE, shown along a plane containing the chain, 
illustrating long-wavelength ferromagnetic domains at $R=0.9\,\mathrm{a}_B$.
}
\label{fig:f3}
\end{figure*}

\subsection{Short-$R$ regime}

\subsubsection{Insulator-to-metal transition}

Correlated electron materials often exhibit metal-insulator
transitions as parameters such as temperature, pressure or crystal
structure are varied \cite{Mott_RMP_1968,Imada_RMP_1998}. 
From the
perspective of the 1D one-band Hubbard model which, as we have seen,
captures the universal aspects of the physics at large $R$ --- and more
generally from the perspective of one-band models with second-order
Umklapp processes, no MIT should occur here \cite{Lieb_PRL_1968}. 
With multiple methods and multiple probes, we show below conclusive
evidence that a MIT occurs in the H chain, and provide a
characterization of the physical origin and properties of the
transition.

The concept of macroscopic localization
\cite{Resta_PRL_1999,Kohn_PR_1964,Sorella_PRB_2011} provides a direct wavefunction-based 
characterization of system properties.  
For periodic systems, one defines the complex polarization 
$Z_N = \langle \Psi_N | e^{i \frac{2\pi}{L} \sum_i \hat{z}_i} | \Psi_N \rangle$, where
$|\Psi_N\rangle$ is the ground state of the $N$ electrons in a supercell of size $L=NR$ along the chain direction. 
The electron localization length $\Lambda = \frac{\sqrt{D}}{2\pi \rho}$ is
related to the complex polarization by $D = - \lim_{N\to \infty} N \log |Z_N|^2$. 
In localized systems, $Z\equiv \lim_{N\to\infty} |Z_N| = 1$, and $\Lambda$ is finite;
in metallic systems $Z = 0$, and $\Lambda \rightarrow \infty$. 

In Fig.~\ref{fig:f3}a, we establish the MIT by computing $Z_N$ with
AFQMC, DMC and VMC. 
(Additional support for a metal insulator transition
is provided by energy gaps as included in the SI.)
For small $R$, all methods give $Z_N$ 
equal to or statistically compatible with $0$ across a wide range of
system sizes $N$, indicating a metallic many-body ground state. For
large $R$, all methods yield a non-zero $Z_N$.  
The many-body methods
point to a transition point located approximately at
$R_{\rm MIT} \sim 1.70(5)$\,$\mathrm{a_B}$.
While there is some uncertainty in the critical value, because of computational limitations,
it is remarkable that two methods working with completely different basis sets and projection algorithms
yield results in quantitative agreement.

In particular the
values of $|Z_N|$   in the insulating phase, which fall further from unity at smaller $R$, 
are sensitive to finite-size effects.
At large $N$, we expect $|Z_N| = 1 - \tilde{g}(\xi/L)$ in the insulating phase,
where $\tilde{g}$ is a scaling function and $\xi$ the MIT correlation length.
The smooth decrease of $|Z_N|$
as $R_{\rm MIT}$ is approached suggests a second-order
transition (related to gap closure). 
We further discuss finite-size corrections
and their implication on $\xi$ in the SI.

Before exploring the origin of the MIT, we briefly discuss magnetic correlations in the metallic phase. 
The evolution of the magnetic moments within DFT 
is shown in Fig.~\ref{fig:f3}b, and the
spin density at $R = 0.9\,\mathrm{a_B}$ is plotted in Fig.~\ref{fig:f3}c as a point of reference. 
To account for spin correlations, the DFT solution breaks translational symmetry to create
antiferromagnetic domains of varying periods, often associated with very diffuse orbitals as seen in Fig.~\ref{fig:f3}c.
Of course, DFT observations are independent-electron 
in nature (and somewhat functional-dependent, see SI).
In the many-body solution, in particular, translational symmetry is restored, 
and two-body correlation functions are needed to probe magnetic correlations. 

\subsubsection{Origin of the MIT and properties of the metallic phase}

\begin{figure}[!ht]
\centering
\includegraphics[width=0.7\columnwidth]{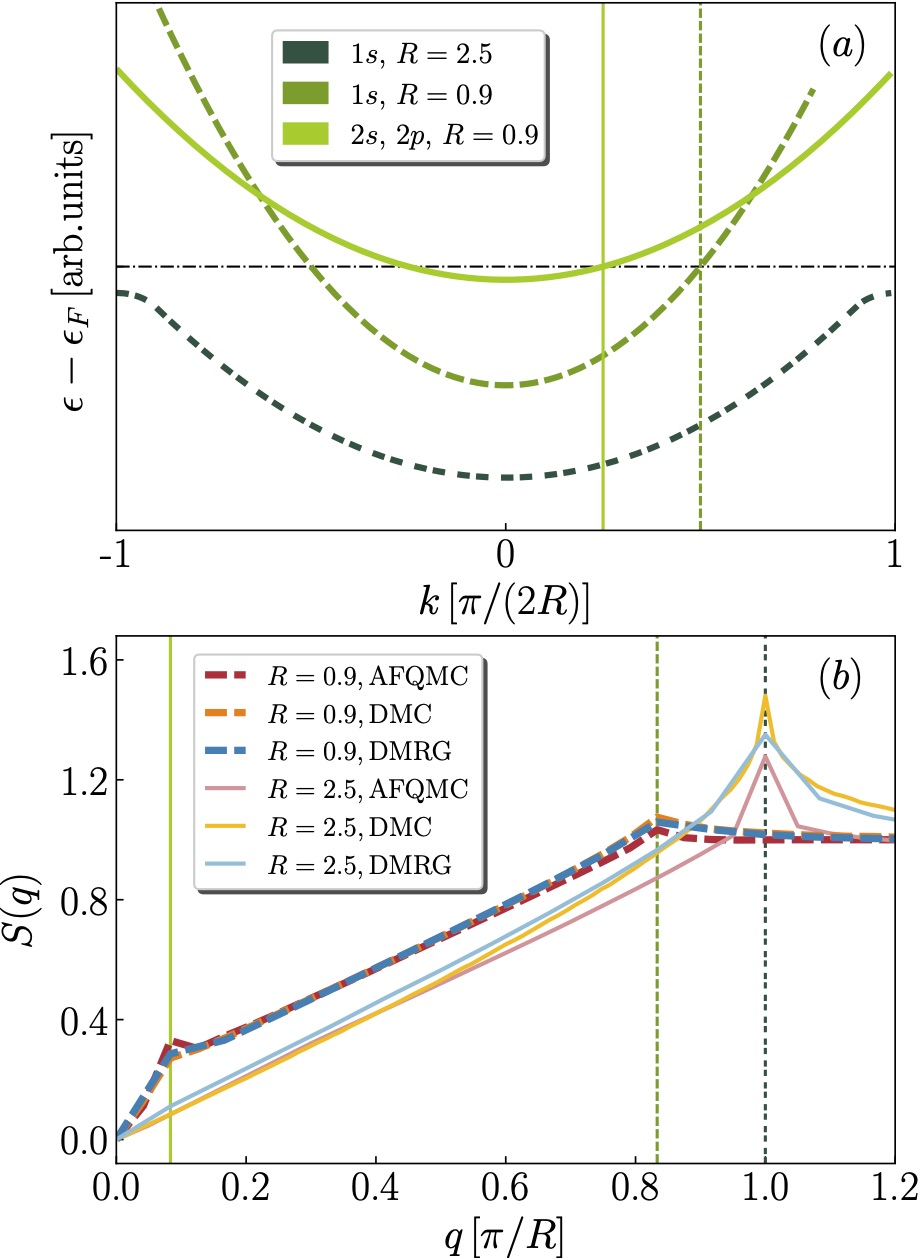}
\caption{
(color online)
{\bf{Mechanism of the insulator-to-metal transition.}}
(a) Schematic illustration of the self-doping mechanism inducing a MIT in the H chain.
Dark green line: large $R$, no band overlap. Light green lines: small $R$, multiple bands crossing $\epsilon_F$.
(b) Structure factor of the spin-spin correlation function at  $R=0.9 \, \mathrm{a}_B$ and at $R=2.5\,\mathrm{a}_B$, 
computed from AFQMC, DMC and DMRG. Wave-vectors $q$ are along the chain.
Vertical lines, as an aid to the eye, mark the kinks in $S(q)$  associated with the Fermi surfaces. 
}
\label{fig:f4}
\end{figure}

It is theoretically established that the ground state of a one-band
model with commensurate filling is insulating. 
Our calculations reveal that the MIT arises from
a self-doping mechanism in which the one-band picture breaks down.  
We illustrate the basic idea in Fig.~\ref{fig:f4}a, using a band-theory based cartoon of the electronic structure.  
The isolated H atom has multiple states, including the occupied $1s$ and excited states $2s$, $2p$, etc. 
At large $R$ the bandwidths are small compared to the energy gaps, and a one-band approximation is reasonable. 
As $R$ is decreased, the bands broaden, 
band overlap occurs, and metallic behavior results. 
Quantitative calculations involve a correlation problem that requires 
treating interactions in a 
multiband situation, effects that are entirely absent in the standard Hubbard model.

We compute 
the spin structure factor,
defined as $S(q) = \frac{1}{N} \, \langle \Psi | \hat{\rho}^\dagger_q \hat{\rho}^{\phantom{\dagger}}_q | \Psi \rangle$, 
where $\hat{\rho}_q = \sum_i \hat{S}_{z,i} e^{i {\bf{q}}\cdot {\hat {\bf r}_i}}$ is the Fourier transform of the spin density at 
${\bf q}=(0,0,q)$, with $\hat{S}_{z,i}$ and 
${\hat {\bf r}_i}=({\hat x_i}, {\hat y_i},{\hat z_i})$ denoting the spin-$z$ and position operators of electron $i$, respectively.
The result
is shown in 
Fig.~\ref{fig:f4}b. 
At $R=2.5$  only one peak is seen at $q=2 k_F^0$, signaling power-law AFM order as discussed earlier.
To probe the nature of the metallic phase, we focus on a representative case of $R=0.9\,\mathrm{a_B}$,
away from the vicinity of the MIT transition.  
$S(q)$ is shown from two different QMC calculations, in supercells of $N=48$ atoms
averaging over 11 twist angles, in addition to DMRG calculations in a
supercell of $N=24$ atoms. In a metallic system we expect peaks at
$q=2k_F$, where $k_F$ is one of the Fermi wavevectors of the system.  
Two cusps are seen in each result at locations, $q_1$ and
$q_2$, in precise agreement among the different calculations.  We
interpret the larger wavevector as arising from the $2 k_F$ process in
the $1s$-dominated lower band. The position  $q_2=2\,(1-x)\,k_F^0$
 then gives  the doping $x$ of this band.
 In a  simple two-band picture,
 the lower wave vector is given by
$q_1=2x\,k_F^0/g$, where $g$ gives the degeneracy of the ``upper                                                         
band'' which is occupied.  At $R=0.9\,\mathrm{a_B}$, the locations of
$q_1$ and $q_2$ satisfy $g \, q_1+q_2 =\pi/R $ in all the results,
with $g=2$. 
This is consistent with a doubly degenerate upper band (e.g. $2 p_{x,y}$),
although multiple bands strongly contribute to the DMRG $S(q)$, as we discuss below.

Within DFT one may characterize the metallic state by the nature of the occupied Kohn-Sham bands. 
Here we use the Perdew-Burke-Ernzerhof (PBE) exchange-correlation functional in DFT calculations.
As $R$ decreases, DFT-PBE solutions indicate a band of mainly atomic $1s$ origin with a relatively large
(but not integer) occupancy and a Fermi level near $k_F^0$, and bands of mostly $2p_x$, $2p_y$ 
character (i.e. $\pi$ bands) becoming occupied, with Fermi levels near $0$. 

\begin{figure}[!ht]
\centering
\includegraphics[width=0.8\columnwidth]{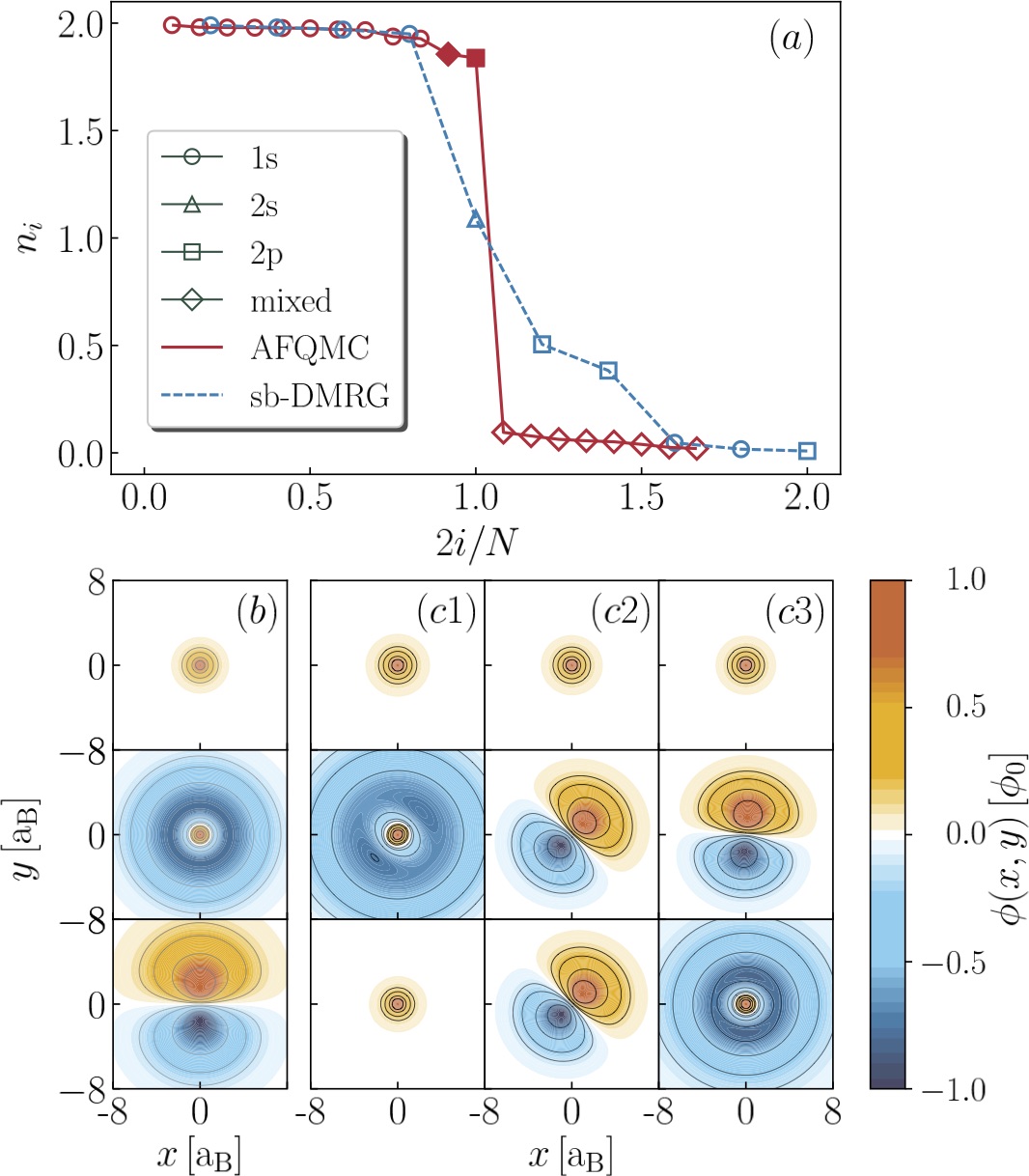}
\caption{
(color online)
{\bf{Insight into the many-body metallic state.}}
(a) Occupancy of NOs,
obtained from the one-body density matrix of the many-body ground state 
at $R=0.9\,\mathrm{a}_B$, from AFQMC (supercell with $N=24$ atoms, twist-averaged) 
and sb-DMRG ($N=10$ atoms with hard-wall boundary). 
The orbital characteristics are indicated, in each dataset, by the marker type.
(b,c) Contour plots illustrating the characters of representative occupied NOs (ordered high to low in $n_i$), 
projected in a transverse plane slicing through an atom, at $R=0.9\,\mathrm{a_B}$ for 
a $\mathrm{H}_{10}$ with OBC and hard-wall boundary, by sb-DMRG (b); 
a $\mathrm{H}_{24}$ supercell from DMRG, at twists $k=0$, $0.1$ and $0.4$, respectively (c1,c2,c3).
}
\label{fig:f5}
\end{figure}

We probe the many-body
metallic state by computing the one-body density matrix, which we then diagonalize.
Each eigenvalue gives the occupancy of the corresponding eigenstate (``natural orbital'', NO).
The occupancies computed by two methods
are shown in Fig.~\ref{fig:f5}a. For the $N=24$ supercell (red symbols),
a sharp transition is seen in orbital occupancy, reminiscent of a Fermi
surface. We label 
the characters of each NO which, similar to Kohn-Sham orbitals, can be analyzed via the single-particle basis or projection onto
atomic orbitals.
In the $N=24$ system the last two
eigenvalues before the sharp drop (11$^{\rm th}$ and 12$^{\rm th}$ points from left, shown as filled symbols) are two-fold
degenerate, consistent with the notion that the metallic phase
corresponds to shifting electrons from $1s$-like to $2p_{x,y}$-like
orbitals. 
However, the many-body results display variations in the
occupancy and the nature of the orbitals, as illustrated in 
Fig.~\ref{fig:f5}b and c. 
We find that the energy differences associated with different occupancies, such as $2s$ 
instead of or together with $2p$, are very small, on the scale of sub-mHa/atom,
and the precise energy ordering is
affected by boundary conditions, supercell sizes, basis sets, as well as the choice of the correlation algorithm. 
This low energy scale is associated with the very diffuse nature of ``upper band'' NOs. 

The occupation and nature of 
these NOs shed
light onto magnetic correlations. 
Electrons associated with the ``lower band'' are 
bound to the chain axis, with a range similar to that of the atom.
$S(q)$ does not show a peak at $q_2$,
but only a kink, indicating short-range correlations from
these electrons.
The electrons in the ``upper bands'' are diffuse, with NOs 
of transverse size $\sim 5 \, \mathrm{a_B}$ (see Fig.~\ref{fig:f5}b,c). 
They display a tendency for magnetic ordering
driven by exchange, as indicated by the kink at $q_1$.  Their average
inter-particle separation along the chain,
$\xi \sim \pi/(2 q_1)$, is large (and becomes larger
as $R_{\rm MIT}$ is approached), with only a small number of electrons
in a large supercell. 
These features make it challenging to determine
the precise behavior of the correlation over a sufficiently long
range, which is further magnified by the sensitivity of the nature of
the upper bands discussed earlier. 
This picture suggests a fascinating 
model of a quasi-one-dimensional electron wire
for probing both magnetic and
charge correlations under the effect of long-range Coulomb
interactions \cite{Schulz1993}.

\section{Conclusions and outlook}

While much has been understood about model systems, most notably the Hubbard model, 
understanding correlated electron physics and reliably
performing predictive computations in real materials 
remain a central challenge in 
condensed matter physics and materials science. 
Significant progress has been made from several fronts, for example with 
the combination of DFT
and the $GW$ and Bethe-Salpeter equation formalisms \cite{GW-BSE-rmp2002,louie2006predicting,C7CS00049A},
approaches based on dynamical mean field theory \cite{Georges_RMP_1996, Marianetti06RMP,MillisNowadnick,ZinglGeorgesSr2RuO4},
as well as several of the methods included here.
In this paper we have focused on pushing state-of-the-art many-body ground-state methods 
towards a comprehensive theory of correlated materials.
Individually, advances are made within each method to address the technical challenges for systematically accurate calculations
of correlated systems in the continuum and thermodynamic limit.
Collectively, with a combination of multiple methods, we are able to achieve a high level of robustness in our
characterization and predictions of the ground-state properties. 

The hydrogen chain provides a good 
compromise between physical/chemical realism and tractability,
We studied
the balance and competition between band structure, electron interaction and multi-orbital physics as a function of a single parameter, the inter-proton 
distance $R$.
At large $R$, the system exhibits a Mott insulating phase with quasi long-range AFM order, and dimerization effects visible in the density, kinetic energy 
and entanglement spectrum. 
Upon decreasing $R$, it transitions via a self-doping mechanism into a metallic phase which  exhibits magnetic correlations. 
At short bondlength the magnetic correlation is modulated by an incommensurate wavevector. 
Near the MIT the occupancy of the ``higher band'' is very low and the size of the magnetic domain grows, 
potentially allowing the possibility to develop global polarization.  
We find that the higher natural orbitals 
(occupied in the metallic phase and important as virtual excitations in the large $R$ insulating phase) have an extremely diffuse and occupation-dependent structure. 
Thus, despite its apparent simplicity, the H chain embodies a surprisingly rich set of the major themes in contemporary quantum materials physics, 
including metal-insulator transitions, magnetism, and intra- and inter-site electron correlation physics.

It is interesting to place our results in  historical context. The study of the MIT was pioneered by Nevill Mott.
In 1949 Mott noted \cite{mott1949basis} that many materials were insulating when band theory considerations indicated that they should be metallic. He argued that one should view the physics in terms of localized atomic-like orbitals   rather than plane waves and observed that in this picture  conductivity required moving an electron from one site to another, creating a particle-hole pair. He argued that this process could be blocked if the charging energy on a given lattice site was large enough to  prevent an additional electron from moving onto it. This physics is famously instantiated in the extensively studied Hubbard model \cite{Hubbard1963}, and our results in the large-$R$ regime are consistent with it.  In subsequent work  \cite{mott1961transition,Mott_RMP_1968} Mott noted that in the insulating phase the unscreened Coulomb interaction would bind the particle and hole to a neutral object, which would not conduct, whereas in a metal the self-consistent screening of the Coulomb interaction would lead to unbound pairs, so that the MIT occurs via a first-order process. Our results demonstrate a different route to the MIT: the H chain becomes metallic via a self-doping process which is (within our resolution) second-order. The carriers in the second band are both very delocalized and at very different $k$-points from the carriers in the dominantly occupied band, so that  excitonic effects are either absent or unobservably weak.  Our work underscores the importance of the interplay of correlation physics with the real materials effects of band crossing and electronic structure.

Experimentally, the physics we have described
could potentially be realized in ultracold atoms in optical lattices, 
arrays of Rydberg atoms \cite{Lukin-Rydberg},
or perhaps with carbon nanotubes, in which carbon chains have been isolated \cite{Shi_nature_2016}. 
Experimental realization would provide an 
{\it{in silico}} insulator-to-metal transition,
which would be a significant contribution to the intensely investigated topic of metallicity in hydrogen \cite{wigner1935possibility,eremets2011conductive,azadi2017nature,Loubeyre2020}.
We also hope that our study will stimulate experimental research on the characterization of hydrogen at extreme density \cite{mao1989optical,loubeyre2002optical,PhysRevLett.114.105305,mazzola2014unexpectedly,Loubeyre2020},
in connection with the theoretical description of astrophysical bodies \cite{guillot2005interiors} 
and the discovery and characterization of exotic magnetic phases in low-dimensional condensed matter systems \cite{dias2017observation,Loubeyre2020}.

More generally, our study establishes the H chain as an important model system 
and demonstrates the power of the synergistic application of independent 
numerical methodologies \cite{LeBlanc_PRX_2015,Motta_PRX_2017,Zheng_Science_2017}.
This mode of attack has the potential to address other challenging problems in condensed matter physics, 
including the systematic investigation of real materials with immediate technological applications by first-principles 
many-body methods.

\section*{Acknowledgments}

We thank T.~Giamarchi, N.~Marzari, A.~Rubio, and M.~van Schilfgaarde for helpful discussions. 
This work was supported by the Simons Foundation as part of the Simons collaboration on the many-electron problem.
The Flatiron Institute is a division of the Simons Foundation.
Computations were carried out on facilities supported by the Scientific Computing Core at the Flatiron Institute (MM, HS)
and by the US Department of Energy, National Energy Research Scientific
Computing Center (ZHC, PH, MM, UR),
on the Pauling cluster at the California Institute of Technology (ZHC, PH, MM, UR)
and on the Storm and SciClone Clusters at the College of William and Mary (FM, MM).
MM acknowledges the IBM Research Cognitive Computing Cluster service for providing resources that have contributed 
to the research results reported within this paper.
GKC acknowledges support from the National Science Foundation under Grant No. OAC 1931258.
FM acknowledges support from the National Natural Science Foundation of China under Grant No. 11674027.
SS and CG acknowledge support from PRIN 2017BZPKSZ and computational resources from CINECA PRACE 2019204934.
SW, EMS, RS and NC acknowledge support from DOE under Grant DE-SC0008696 and Swiss National Science Foundation.

\end{document}